\documentstyle[12pt]{article}
\begin{document}
\title{ Paragrassmann Analysis and Quantum Groups}
\author{\em A.T.Filippov\thanks{present address: Dip. di Fisica Teorica,
Univ. di Torino, via P.Giuria 1, I-10125 TORINO}~\thanks{e-mail address:
FILIPPOV@TORINO.INFN.IT~~39163::FILIPPOV}~, A.P.Isaev\thanks{e-mail address:
ISAEVAP@THEOR.JINRC.DUBNA.SU} and A.B.Kurdikov \\
Laboratory of Theoretical Physics, JINR, Dubna \\
 SU-101 000 Moscow, Russia }
\date{}
\maketitle

\begin{abstract}
Paragrassmann algebras with one and many paragrassmann variables
are considered from the algebraic point of view
without using the Green ansatz. A differential operator
with respect to paragrassmann variable
and a covariant para-super-derivative are
introduced giving a natural generalization of the Grassmann
calculus to a paragrassmann one.
Deep relations between paragrassmann algebras and quantum groups
with deformation parameters being roots of unity are
established.

\end{abstract}
\newpage

\section{Introduction}

Paragrassmann algebras (PGA) are interesting for several reasons. First, they
are relevant to conformal field theories [1]. Second, studies of anyons
and of topological field theories show the necessity of unusual statistics
[2], [3]. These include not only the well-known Green-Volkov
parastatistics [4] but fractional statistics as well.
There are also some hints (e.g. Ref.[5]) that PGA have a connection to
quantum groups. Finally, it looks aesthetically appealing to find a
generalization of the Grassmann analysis [6] that proved to be so successful
in describing supersymmetry.

Recently, some applications of PGA have been discussed in literature.
In Ref.[7], a parasupersymmetric generalization of quantum
mechanics had been proposed. Ref.[8]
has attempted at a more systematic consideration
of the algebraic aspects of PGA based on the Green ansatz [4] and introduced,
in that frame, a sort of paragrassmann
generalization of the conformal algebra.
Applications to the relativistic theory of the first-quantized spinning
particles have been discussed in Ref.[9]. Further references can be found
in [1],[5],[7],[8].

The aim of this paper is to construct a consistent generalization of the
Grassmann algebra (GA) to a paragrassmann one preserving, as much as possible,
those features of GA that were useful
in physics applications. The crucial point
of our approach is defining a generalized derivative in paragrassmann
variables. This is shown to relate PGA, in a
natural way, to $q$-deformed algebras
and quantum groups with $q$ being a root of unity.
In this paper, we mainly concentrate on the algebraic aspects leaving the
applications to future publications. It should be stressed that we do not
use the Green ansatz although natural matrix realizations of the
algebraic constructions are given.

Section 2 treats the algebra generated
by one paragrassmann variable $\theta $,
$\theta^{p+1} =0$, and automorphisms of this
algebra. In Section 3, a notion of
generalized differentiation is introduced and discussed. It uses special
automorphisms preserving a natural grading of PGA
and naturally moves into action
the roots of unity, $q\; \;(q^{p+1} =1)$. The generalized differentiation
coincides with the Grassmann one for
$p=1$, and with the standard differentiation
when $p\rightarrow \infty $. For intermediate cases $1<p<\infty $, the
structure of the algebra depends on
the arithmetic nature of its order $p+1$.
This is briefly discussed in Section 4 where
the simplest PGA with many variables
$\theta_{i}$ are defined (PGA with $N$ variables will be denoted
as $\Gamma_{p}(N)$). They satisfy the nilpotency condition
$\theta^{p+1}=0$ where $\theta$ is any linear combination of $\theta_{i}$,
and appear to be naturally related to the
non-commutative spaces satisfying the commutation relations
$\theta_{i}  \theta_{j}=q \theta_{j} \theta_{i}\;,\; i<j$.
These and other relations presented in this
paper demonstrate a deep connection between PGA and quantum groups with
deformation parameters $q$ being roots of unity. Two of
the most obvious are presented in the Sections~4 and 5.

\section{Paragrassmann Algebra with One Variable}

We start by defining the paragrassmann algebra $\Gamma_{p}(1)$ (or, simply
$\Gamma $), generated by one nilpotent variable $\theta $
($\theta^{p+1}=0 ,\;p$ is some positive integer). Any element of the algebra,
 $a\in \Gamma$, is a polynomial in $\theta$ of the degree $p$,
 \begin{equation}
 \label{aa}
    a=a_{0}+a_{1} \theta+\cdots +a_{p} \theta^{p}\; ,
 \end{equation}
where $a_{i}$ are real or complex numbers or, more generally, elements of
some commutative ring (say, a ring of complex functions) [10].
It is useful to have a matrix realization of this algebra. One may
regard $a_i$ as coordinates of the vector $a$ in the basis
($1,~ \theta ,~ \dots ,~ \theta^{p}$).
Defining the operator of multiplication by $ \theta$,
          \begin{equation}
 \theta ( a)=a_{0} \theta+\cdots +a_{p-1} \theta^{p} ,
          \end{equation}
we see that it can be represented by the triangular $(p+1)\times
(p+1)$-matrix acting on the coordinates of the vector $a$:
			 \begin{equation}
			 \label{thmn}
 ( \theta)_{mn} = \delta_{m,n+1} ~,~
  (\theta^{k})_{mn} = \delta_{m,n+k} ~,
			 \end{equation}
 $m,n=0,1,\dots ,p$. We may now treat elements of the algebra
as matrices. In view of Eq.(3), any element $a\in\Gamma$ can be represented
by the matrix
			 \begin{equation}
	(a)_{mn}=\left\{ \begin{array}{ll} a_{m-n} & {\rm if} \; m\geq n,\\

                                               0 & {\rm if} \; m<n.
					\end{array} \right.
				 \end{equation}
This matrix representation of the algebra is obviously an isomorphism.

A very important construction related to the algebra $\Gamma$ is its group
of automorphisms consisting of the linear maps $a\rightarrow
g(a)$ that preserve the multiplication,

					 \begin{equation}
		g(\alpha a+\beta b)=\alpha g(a)+\beta g(b)\; ,
					 \end{equation}
					 \begin{equation}
					 \label{a}
					 g(ab)=g(a)g(b),
					 \end{equation}
where $\alpha,\ \beta $ are numbers. It is clear that any automorphism
is defined by $p$ parameters $\gamma_{m}\;,m=0\dots p-1$:
					 \begin{equation}
					    \label{g}
		 g(\theta )=\sum_{m=0}^{p-1} \gamma_{m}\theta^{m+1},
					 \end{equation}
or, in the infinitesimal form
					 \begin{equation}
 \delta_{\epsilon} \theta=\sum_{m=0}^{p-1} \epsilon_{m}\theta^{m+1} .
			 \end{equation}
Omitting the obvious summation symbols we have
\[  \delta_{\epsilon} a \equiv \delta_{\epsilon} (a_{k}\cdot \theta^{k})=
a_{k}\cdot \delta_{\epsilon} \theta^{k}=
a_{k}\cdot k\epsilon_{m}\theta^{k+m}. \]
Rewriting this as the infinitesimal transformation of the coordinates
 \[  \delta_{\epsilon} a_{k}\equiv \epsilon_{m}(G^{m})_{kl} a_{l} \; ,  \]
we find the matrix elements of the Lie--algebra generators $G^{m}$
\begin{equation}
	(G^{m})_{kl}=l\delta_{k-l,m}  \; ,
\end{equation}
and the commutation relations
\begin{equation}
[G^{m},G^{n}]=(n-m)G^{m+n} ,
\end{equation}
where $G^{m+n}=0$, if  $m+n\geq p$. Being the generators of the automorphism
group, $G^{m}$ define differentiations of the algebra $\Gamma$, the
classical ones satisfying the Leibniz rule.
However, it is impossible to treat any of
them as a differentiation with respect to $\theta$. In fact,
\[  G^{m} (\theta^{n})=\left\{
\begin{array}{ll}n\theta^{m+n} &
{\rm if} \; n+m\leq p,  \\
0 & {\rm if} \; n+m>p,
 \end{array} \right. \]
but we would rather expect a differentiation
$\partial \equiv \partial /\partial\theta$
to act as
\begin{equation}
\label{nor}
 \partial (1)=0,\ \partial(\theta) =1,\
\partial(\theta^{n})\propto \theta^{n-1}\; ,\
n>1.
\end{equation}
It is easy to see that the condition $\partial(\theta)=1$
together with the standard
Leibniz rule,
$\partial (ab)= \partial ( a) \cdot b + a \cdot \partial (b)$,
completely define the action of $\partial$ on any $a\in \Gamma$,
 but this immediately leads to a contradiction
\[ 0\equiv \partial(\theta^{p+1})=
{\rm (via\; Leibniz\; rule)\ }=(p+1)\theta^{p} .
 \]
 This is a manifestation of the general fact about nilpotent algebras known
 even for the Grassmann case: once the conditions of the type
  (\ref{nor}) are required, the Leibniz rule is to be deformed.

 \section{Generalized Differentiation}

To introduce a useful definition of $\partial $ we suggest a generalized
	Leibniz rule ($g$-Leibniz rule)
	\begin{equation}
  \label{l}
	\partial(ab)=\partial (a)\cdot b+g(a)\cdot \partial (b) \;,
	\end{equation}
where $g$ is some automorphism of the algebra $\Gamma_{p}$. For the
Grassmann case $(p=1)$ we have $g(\theta)=-\theta$, so that
$\gamma_{0}=-1$. This is usually written as
$g(a)=(-1)^{(a)}a $ where (a) is the Grassmann parity of the element $a$.
For arbitrary $p$, the automorphism $g$ and, hence, the derivative
$\partial\ $ are completely
fixed by the conditions $\partial(\theta)=1$ and
 $\partial(\theta^{2})\propto\theta$. These, by (\ref{l}) and (\ref{g}), give
\[ \gamma_{m}=0\ {\rm for\ }m>0\;, \; g(\theta)=\gamma_{0}\theta \;, \]
 \[ \partial (1)=0\;,\; \partial(\theta^{n})=(1+\gamma_{0}+\cdots
 +\gamma^{n-1}_{0})\theta^{n-1} \;,  \]
  and from $\partial(\theta^{p+1})\equiv 0$ we get
 \begin{equation} 1+\gamma_{0}+\cdots +\gamma^{p}_{0}\equiv \frac
	 {1-\gamma^{p+1}_{0}}{1-\gamma_{0}}=0 \;,
	  \end{equation}
	 so that $\gamma_{0}$ is fixed to be a root of unity. For the moment,
we choose $\gamma_{0}$ to be the \it prime \rm root, i.e.
	 \begin{equation}
	 \gamma_{0}=q\equiv e^{2\pi i/(p+1)} = (-1)^{2/(p+1)}.
	 \end{equation}
	By introducing the notation
	 \begin{equation}
	 \label{nn}
	 (n)_{q}\equiv 1+q+\cdots +q^{n-1}=\frac {1-q^{n} }{1-q}\;,
	 \end{equation}
	 the action of $\partial $ can be written as
	 \begin{equation}
	 \label{act}
	 \partial (\theta^{n})=(n)_{q}\theta^{n-1},
	 \end{equation}
 and so the matrix elements of $\partial$ in the basis $\{\theta^{m}\},
						m=0,\dots ,p\ $, are
	 \begin{equation}
	 (\partial)_{mn}=(m+1)_{q} \delta_{m+1,n}\;.
	 \end{equation}
Since $(p+1)_{q}=0$, the operator $\partial$ is nilpotent, $\partial^{p+1}=0$.
It is not hard to see that the operators $\partial$ and $\theta$ satisfy the
$q$-deformed commutation relation
	 \begin{equation}
	 \label{com}
	 [\partial ,\theta ]_{q}\equiv \partial \theta-q\theta \partial =1\;.
	 \end{equation}

	 The Grassmann case for $p=1$ and the classical one in the limit
	  $p\rightarrow \infty$ are evidently reproduced.
The last equation is suggestive of a relation between PGA and much discussed
$q$-deformed oscillators and quantum groups
(see, e.g. Refs.[12] --- [14], [17], [18]) with the deformation
parameter $q$ being a root of unity. We will return to this point at
the end of the paper.

Consider now the algebra $\Pi_{p}(1)$ (or, simply $\Pi$) generated by both
$\theta$ and $\partial$. Since Eq.(\ref{com}) makes it possible to push all
$\partial$'s to the right of $\theta$'s, the complete basis of $\Pi$ might be
given by $(p+1)^{2}$ monomials $\{\theta^{m}\partial^{n}\},\ m,n=0,\dots ,p$.
(Their linear independence is quite evident in the matrix representation).
Thus $\Pi$ is isomorphic, as an associative algebra, to the general matrix
algebra of the order $p+1$ with the natural
``along-diagonal'' grading
\begin{equation}
deg(\theta^{m}\partial^{n})=m-n         \;.
\end{equation}
Note that this grading makes it possible to rewrite the $g$-Leibniz rule
(\ref{l}) in a complete visual correspondence to the Grassmann case
\begin{equation}
\partial (ab)=(\partial a)b+(-1)^{\frac {2}{p+1} deg \; a} a(\partial b)
\end{equation}
(one can interpret the quantity $(a)=\frac {2}{p+1} deg \; a$
as the paragrassmann parity of the element $a$)

Note also that since the automorphisms of $\Gamma$ can be
represented by $(p+1)$-matrices, they must have an expression in terms of
$\theta$ and $\partial$. In particular, the operator $g$ from Eq.(\ref{l})
is expressed as
\begin{equation}
\label{ggg}
g=\partial \theta -\theta \partial =1+(q-1)\theta \partial .
\end{equation}
Its matrix elements are
\begin{equation}
(g)_{mn}=q^{m}\delta_{mn}
\end{equation}
In the mathematical literature (see, e.g. Ref.[11]), generalized
differentiations satisfying the $g$-Leibniz rule
(\ref{l})  are called $g$-differentiations. Mathematicians also consider a
further generalization, called $(g,\bar{g})$-differentiation, which
satisfies the rule
\begin{equation}
\label{yy}
\partial (ab)=\partial(a)\cdot \bar{g}(b)+g(a)\cdot \partial (b)\;.
\end{equation}
This generalization of the Leibniz rule is related to
a special representation of the algebra
$\Gamma$ by $2\times 2$-matrices with elements in $\Gamma$
\begin{equation}
a\longmapsto \left(
	 \begin{array}{cc}
	 g(a) & \partial (a) \\
	 0    & \bar{g}(a)
	 \end{array}
	 \right) \equiv M(a)\;.
\end{equation}
If $g$ and $\bar{g}$ are algebraic homomorphisms, i.e. satisfying
Eq.~(\ref{a}),
then Eq.~(\ref{yy}) is equivalent to the homomorphism condition
\[ M(ab)=M(a)M(b)\;.  \]
All this is obviously applicable to the $g$-Leibniz rule and to the standard
one as well.

For physical applications, it seems more reasonable to use for $g$ and
$\bar{g}$
some automorphisms rather than just homomorphisms. Although we think that
Eq.(\ref{l}) looks more natural than Eq.(\ref{yy}), the latter can be used
to define a ``real''  differentiation, i.e. having real matrix elements.
In fact, choosing for $g$ and $\bar{g}$ the automorphisms defined by
\begin{equation}
\label{isa}
g(\theta )=q^{1/2}\theta \;,\;
\bar{g}(\theta )=q^{-1/2}\theta  \;,
\end{equation}
we find that
\begin{equation}
\label{ddd}
\partial (\theta^{n})=[n]_{\sqrt {q}}\theta ^{n-1}\ ,
\end{equation}
with the popular notation
\begin{equation}
[n]_{\sqrt {q}}\equiv \frac {q^{n/2}-q^{-n/2}}{q^{1/2}-q^{-1/2}}
 =q^{(1-n)/2} (n)_{q} \; .
 \end{equation}
 This is obviously a real number. The operators $g$ and $\bar{g}$ have
the matrix elements
 \begin{equation}
 \label{gmn}
 (g)_{mn}=q^{m/2}\delta_{mn} \;,\;
 (\bar{g})_{mn}=q^{-m/2}\delta_{mn} \;,
 \end{equation}
 and the following expression in terms of $\theta$ and $\partial$
 \begin{equation}
\label{is}
 g=\partial \theta- q^{-1/2} \theta \partial \; ,\; \;
 \bar{g}=\partial \theta- q^{1/2} \theta \partial  \;.
 \end{equation}
The first equation in (\ref{is}) is an analog of Eq. (\ref{ggg})
while the second one may be considered as an analog of (\ref{com}).
One can easily recognize in formulas (\ref{is}) the definition
of the $q$-deformed quantum oscillator (see, e.g. [12], [17], [18]). We
 will exploit this variant of differentiation in the last section of
this paper.

In addition to the $g$-differentiation, one can also construct an
inverse operation, or $g$-integration,
$(\partial )^{-1}=\int_{\theta}$. To do that, one has to ``regularize''
$\theta$ and $\partial$ by introducing a formal parameter dependence to
$\theta$ and $(n)_{q}\;$, e.g. $\theta_{\epsilon }=\theta +\epsilon^{2}\ ,\
q_{\epsilon }=q^{1+\epsilon } $. Then, the following simple definition
\[  \int_{\theta} \theta^{n}_{\epsilon }=\frac{\theta^{n+1}_{\epsilon}}
{(n+1)_{q_{\epsilon }}} , \]
makes sense and one can check that
\[  \partial \int_{\theta} =1   \]
in the limit $\epsilon \rightarrow 0$. This definition satisfies
the $g$-modified partial integration rule
\[  \int_{\theta}(\partial a)b=ab-\int_{\theta}g(a)\partial b\ .  \]
In the limit $p \rightarrow \infty$ this definition reproduces the usual
indefinite integral. Our definition of the $\theta$-integration has
no relation to the standard Grassmann integration.
A possible definition of the integration over $\theta$ that generalizes
the Grassmann integration to the paragrassmann one has earlier been
addressed in Ref.[15].

Up to now, we have been discussing the paragrassmann algebras
with complex (or real) coefficients $a_{n}$.
In some applications (e.g. in constructing parasupersymmetries)
one has to deal with $a_{n}$ (Eq.(\ref{aa}))
belonging to a wider commutative ring, for instance, the ring
of the differentiable functions of a real or complex variable $t$, i.e.
$a_{n}=a_{n}(t)$.
For such an algebra, it is possible to define a sort of
``para--super--covariant derivative''
\begin{equation}
\label{cd}
D=\partial_{\theta } +\frac{1}{(p)_ {q}!}\theta^{p}\partial_{t}\ ,
\end{equation}
where $\partial_{\theta} \equiv \partial $ and the standard notation is used
\begin{equation}
\label{gf}
(n)_{q}! = (n)_{q}(n-1)_{q} \cdots (1)_{q}   \;.    
\end{equation}
This derivative obviously satisfies the $g$-Leibniz rule (\ref{l}) and
 may be considered as a root of $\partial_{t}\; $ since
\begin{equation}
D^{p+1}a(t;\theta )=\partial_{t} a(t;\theta )  \;.
\end{equation}
Unlike $\partial_{\theta }$, the derivative $D$ possesses
 eigenfunctions, the $q$-exponentials
\[  e_{q}(t;\theta )=e^{t} \sum_{n=0}^{p} \frac{\theta^{n}}{(n)_{q}!}\;,  \]
\[  D e_{q}(\lambda^{p+1}t;\lambda \theta)\;
        =\lambda e_{q}(\lambda^{p+1}t;\lambda \theta).  \]
In the limit $p\rightarrow \infty$ we have $e_{q}(t;\theta )\rightarrow
{\rm exp}(t+ \theta )$.

\section{Many Paragrassmann Variables }

Our discussion of the paragrassmann algebras $\Gamma_{p}(1)$ and $\Pi_{p}(1)$
was completely
general and did not rely on special
matrix representations for $\theta$ and
$\partial$. In fact, different representations could be classified if we
relaxed our assumption for $q$ to be the prime root of unity, $q_{p}=
{\rm exp}(2\pi i/(p+1))$. One would
find that the structure of the algebra
$\Pi_{p}(1)$ depended on the arithmetic properties of $(p+1)$. The
simplest case is when $(p+1)$ is a prime integer. Then the multiplicative
group of roots of unity, ${\bf Z}_{p+1}$,
has no subgroups; any root generates the whole group and may be used for
defining $\partial$.
If $p+1$ is a composite number having divisors $p_{i}$,
the group of roots contains subgroups,
${\bf Z}_{p_{i}}$,
generated by the roots $q_{i}={\rm exp}(2\pi i/p_{i})$.
Correspondingly, the algebra $\Gamma_{p}(1)$ has the subalgebras generated
by $\theta^{p_{i}}$ having the following property: if we define $\partial$,
with $q$ in Eq.(\ref{act}) replaced by $q_{i}$, we will find that
$\partial \equiv 0$ over all subalgebra generated by $\theta^{p_{i}}$.
It follows that
we can choose $q$ only of the \it primitive \rm roots, i.e. those
that generate the entire group
${\bf Z}_{p+1}$ (not just a subgroup).

In summary, when
$(p+1)$ is a prime number, any root is primitive (except unity)
and, hence, there are $p$ possibilities to define $\partial$.
For a composite $(p+1)$, the number of possible differentiations is
equal to $\phi(p+1)$ which is the number of positive integers smaller
than $(p+1)$
and relatively prime to it. Such an ambiguity becomes crucial
when we turn to the many-$\theta$ case
giving rise to the existence of a series
of nonequivalent paragrassmann algebras $\Gamma_{p}(N)$. Needless to say,
it is a pure $p>1$ effect.

Leaving these subtleties to some further paper we present here
just the simplest inductive construction of $\Gamma_{p}(N)$.
Starting with $N=2$, define
\begin{equation}
\label{tt}
\theta_{1}=g\otimes \theta\ ,\ \theta_{2}=\theta \otimes 1 \;,
\end{equation}
where $\theta$ and $g$ have been defined in the previous section.
It is easy to see that
\begin{equation}
\theta_{1}\theta_{2}=q\theta_{2}\theta_{1}\ ,\ \theta^{p+1}_{i}=0\ .
\end{equation}
The crucial fact is that the definition (\ref{tt}) allows for nilpotency of any
linear combination of $\theta_{1}$ and $\theta_{2}$. In fact, as one can
easily derive by induction,
\begin{equation}
\label{oo}
(a_{1}\theta_{1}+a_{2}\theta_{2})^{n}=\sum_{k=0}^{n}{n\choose k}_{q}
	 a^{k}_{1} a^{n-k}_{2}\theta^{n-k}_{2}\theta^{k}_{1}\ ,
\end{equation}
where
\begin{equation}
\label{oio}
{n\choose k}_{q}=\frac{(n)_{q}!}{(k)_{q}!(n-k)_{q}!}
\end{equation}
are $q$-deformed binomial coefficients, the polynomials in $q$
(the Gauss polynomials). Remembering now the definitions (\ref{gf}) and
 (\ref{nn}), we immediately prove that
 \begin{equation}
 (a_{1}\theta_{1}+a_{2}\theta_{2})^{p+1}=0\; ,
  \end{equation}
 as long as $q$ is a primitive root of unity.

Suppose now that we have constructed  the algebra $\Gamma_{p}(N)$ satisfying
the relations
\begin{equation}
 \label{N}
\theta_{i}\theta_{j}=q\theta_{j}\theta_{i}\; ,\ i<j\; ,\; i,j=1,\dots ,N
\; ,\;
\end{equation}
\begin{equation}
\label{Na}
(\sum^{N}_{i=1} a_{i}\theta_{i})^{p+1}=0\; .
\end{equation}
Then, $N+1$ matrices $\vartheta_{i}$ satisfying (\ref{N}) and (\ref{Na})
 can be constructed in analogy to Eq.(\ref{tt})
 \begin{equation}
 \label{isi}
\vartheta_{i}=g\otimes \theta_{i}\; ,\ i=1,\dots ,N, \;\vartheta_{N+1}=\theta
\otimes 1 .
\end{equation}
The proof of the identity (\ref{Na}) is performed in full analogy
with the $N=2$ case.
Thus, the induction ensures the existence of the algebras $\Gamma_{p}(N)$
satisfying the conditions
 (\ref{N}) for all $N$. As has been noted above, it is a simplest
construction of the paragrassmann algebra with many generators.
The complete
classification of all admissible forms of $\Gamma_{p}(N)$ is an
interesting but a separate problem.

It is rather amusing that the consideration of paragrassmann algebras
naturally leads to the objects introduced in the context of quantum groups.
In fact, the generators of the algebra $\Gamma_{p}(N)$, determined by the
relations of the type (\ref{N}) and (\ref{Na}), might be considered as
coordinates of a certain nilpotent
quantum hyperplane similar to those of Refs.[13], [14]. Such an object and,
especially, its $\partial$-extensions (defined by its
automorphisms) look rather interesting both from algebraic
and from quantum-geometric [16] points of view.
Here we just briefly outline problems arising in this area.

Let us consider an algebra $\Gamma_{p}(N)$ with the commutation relations
\begin{equation}
\label{i}
\theta_{i}\theta_{j}=q^{\rho_{ij}}\theta_{j}\theta_{i}\;,\;i,j=1,\dots ,N \;,
\end{equation}
where $q$ denotes the prime root of unity.
The requirement for $q^{\rho_{ij}}$ to be a primitive root is equivalent
to the requirement for $\rho_{ij}$ to be invertible elements of the ring
${\bf Z}_{p+1}$. Then, let us define differentiations $\partial_{i}$
satisfying the normalization conditions,
\begin{equation}
\label{iii}
\partial_{i}(\theta_{k})=\delta_{ik}
\end{equation}
and the $g$-Leibniz rule
\begin{equation}
\label{iv}
\partial_{i}(ab)=\partial_{i} (a)\cdot b +g_{i}(a)\cdot \partial_{i} (b) \;
\end{equation}
where the action of the automorphisms $g_{i}$ on $\theta_{k}$ is
A
\begin{equation}
\label{v}
g_{i}(\theta_{k})=q^{\nu_{ik}}\theta_{k} \;.
\end{equation}
These conditions determine the commutation relations in the operator form
\begin{equation}
\label{vi}
\partial_{i}\theta_{k}=\delta_{ik} +q^{\nu_{ik}}\theta_{k}\partial_{i} \;.
\end{equation}
It is not difficult to show that
\begin{equation}
\label{vii}
 \partial_{i}\partial_{j} = q^{\rho_{ij}} \partial_{j}\partial_{i}
 \;,
\end{equation}
and for $i \not= k$,
\begin{equation}
\label{viii}
  \nu_{ik} =\rho_{ki}=-\rho_{ik}\;,
		\end{equation}
while the diagonal elements $\nu_{ii}$ remain not specified. There were
no problems so far. But adding the requirement that any linear
combination of $\partial_{i}$ must also be a differentiation satisfying
(\ref{iv}) with certain $\tilde{g}$ immediately gives
\begin{equation}
\label{ix}
g_{i}(a)=g_{j}(a)=\tilde{g}(a)
\end{equation}
and, therefore,
\begin{equation}
\label{x}
\nu_{ik}=\nu_{jk}          \;.
\end{equation}
The conditions (\ref{viii}) and (\ref{x}) are in general hard to be satisfied
together. For $N=2$ the solution exists
\begin{equation}
\label{xi}
\nu_{11}=\nu_{21}=-\nu_{12}=-\nu_{22}=({\rm some\;invertible\;element\;of\;}
{\bf Z}_{p+1}\;).
\end{equation}
Eqs.(47),(49),(50) then give the commutation relations (41),(45),(46)
formally coinciding with one of formulations of the quantum plane
(up to hermiticity properties)
$\theta_{1}\theta_{2}=q\theta_{2}\theta_{1}\;, \;
 \partial_{1}\partial_{2} = q \partial_{2}\partial_{1}\;, \;
 \partial_{i}\theta_{i}=1 +q^{\nu_{ii}}\theta_{i}\partial_{i} \; $.
But for $N>2$ the equation (\ref{N})
ensures the existence of the algebra (\ref{i}) with all $\rho_{ij}=1$ for
$i<j$, which is evidently inconsistent with (47),(\ref{x})
(unless $p=1$, of course).
This demonstrates the necessity of a more detailed consideration
of the algebra $\Gamma_{p}(N)$ and of its automorphisms which is not
attempted in this paper.

It is however possible to construct another interesting extension of
$\Gamma_{p}(N)$ for arbitrary $N$ by introducing a different definition
for the derivatives $\partial_{i}$ (we will see in a moment that
$p$ must be even). Forgetting about the Leibniz rule, it is most
natural to define operators $\partial_{i}$ by
the inductive procedure similar to (\ref{isi}):
\begin{equation}
\label{GGG}
\tilde \partial_{i}=g\otimes \partial_{i},\;\;i=1,\dots ,N,\;\;\;
\tilde \partial_{N+1}=\partial \otimes 1 \; ,
\end{equation}
where we have also slightly modified the definition of $\partial$ ($g$ is
defined by Eq.(22)):
\begin{equation}
\partial \theta - q^{2} \theta \partial = 1 \; ,
\label{HHH}
\end{equation}
\begin{equation}
\partial \theta -  \theta \partial = g^{2} \; .
\label{hHH}
\end{equation}
 From these equations and from definitions of $\theta_{i}$ and
$\partial_{i}$ ($i=1,\dots ,N$) we obtain the following algebra
\begin{eqnarray}
\theta_{i}\theta_{j} & = & q \theta_{j}\theta_{i} \;\;\;\;
i<j \; ,\nonumber \\
\partial_{i}\partial_{j} & = & q^{-1}\partial_{j}\partial_{i}
\;\;\;\;i<j  \;  , \nonumber \\
 \partial_{i} \theta_{j} & = & q \theta_{j} \partial_{i} \; \;\;\;
 i \neq j \; ,
\nonumber \\
\partial_{i}\theta_{i} & - & q^{2} \theta_{i}\partial_{i} =
1+(q^{2}-1)\sum_{k>i} \theta_{k}\partial_{k}   \; .
\end{eqnarray}
These are the formulas for differential calculus on the quantum
hyperplane constructed by J.Wess and B.Zumino [16].
{}From their results it follws that, instead of our $g$-Leibniz rule
(\ref{iv}), the derivatives $\partial_{i}$ satisfy
$$
\partial_{i}(ab)=\partial_{i} (a)\cdot b +
g^{j}_{i}(a)\cdot \partial_{j} (b) \; .
$$
Note that nilpotency of the linear combinations $a_{i}\theta_{i}$ and
$b_{i}\partial_{i}$ as well as nondegeneracy of $\partial$
(see Eq.(\ref{HHH})) are guaranteed if both $q$ and $q^{2}$ are
primitive roots of unity. This requires that $p$ is even integer.

These formulas may also be interpreted as the definition
of the covariant $q$-oscillators [18]
or, else, as the central extension of the quantum symplectic space relations
for the quantum group $Sp_{q}(2N)$ (see L.D.Faddeev et al. [13]).
So, this example dramatically demonstrates a deep relation between
paragrassmann algebras and quantum groups.
Another example will be presented in the next section.

\section{Discussion}

In this paper, we have introduced the basic ideas of a rather general approach
to constructing paragrassmann algebras with differentiations. One may ask
 a question: what are the relations an algebra must satisfy to be
called paragrassmann? In fact, one of them is clear -- it is the $p$-nilpotency
of any linear combination of generators $\theta_{i}\;(i=1,\dots ,N)$ or,
equivalently,
\begin{equation}
\label{j}
\sum_{\sigma \in S_{p+1}}\theta_{\sigma (i_{0})}\theta_{\sigma(i_{1})}\cdots
\theta_{\sigma(i_{p})}=0\;,
\end{equation}
where the sum is taken over all permutations of the indices. It is clear that
the algebra with the only identity (\ref{j}) would be very hard to handle. So,
one must impose some additional restrictions. A variant of those, known as
the Green ansatz (see Ref.[4]), consists in
taking each paragrassmann generator
$\theta_{i}$ to be a sum of $p$ mutually commuting Grassmann numbers.
In addition to Eq.(\ref{j}), this gives the condition
\begin{equation}
\label{jj}
[[\theta_{i_{1}},\theta_{i_{2}}],\theta_{i_{3}}]=0 .
\end{equation}
 Such an algebra admits a sort of analysis (see [4])
which unfortunately quickly becomes messy as $p$ increases.

As has been shown above, using the conditions (\ref{i}) instead of (\ref{jj})
(with certain restrictions on $\rho_{ij}$ coming from (\ref{j}))
gives a much simpler algebra possessing matrix
representation, differentiations and, as we might suspect, many other useful
properties analogous to its Grassmann ancestor. These are the algebras we
should call paragrassmann. One can easily check that the conditions (\ref{i})
and (\ref{jj}) are not particular cases of each other, and so the algebras
$\Gamma_{p}(N)$ of the present paper are different from those
of Ref.[4].

The most curious is the connection between paragrassmann and $q$-deformed
 algebras. In fact, our interest to paragrassmann algebras was initiated by
searching for the parafermionic extensions of the Virasoro algebra (which we
are going to present in the next paper). So, coming into play of roots of
unity, $q$-oscillators, etc. was somewhat surprising.
To make this connection more apparent, we give here a representation
of the $q$-deformed algebra
$U_{q}(su(1,1))$ in terms of the paragrassmann variable $\theta$
and $(g,\bar{g})$-differentiation $\partial$
(the analogous
construction for $U_{q}(su(1,1))$ from the q-deformed oscillator
was considered in Ref.[17]).
This can be done by representing the homomorphisms $g$ and $\bar{g}$
 from (\ref{yy}) as operators which are inverse to each other (see Eq.
(\ref{isa}))
\begin{equation}
g=q^{\tilde{N}} \; , \; \; \bar{g}=q^{-\tilde{N}} \; .
\end{equation}
Then, defining the generators $N,\;E_{+}$ and $E_{-}\; $
$$
N=\tilde{N}+1/2 \; ,
$$
	\begin{equation}
	\label{isae}
E_{+}=\frac{1}{(q^{1/2}+q^{-1/2})^{1/2}} \theta^{2} \; ,
\end{equation}
$$
E_{-}=\frac{1}{(q^{1/2}+q^{-1/2})^{1/2}} \partial^{2} \; ,
$$
and using Eq.(\ref{is}),
it is not hard to check that generators (\ref{isae}) satisfy
 the well-known relations of the quantum algebra $U_{q}(su(1,1))$
in the Drinfeld-Jimbo form
$$
[N,E_{\pm }]=\pm E_{\pm } \;,
$$
$$
[E_{+},E_{-}]=-[ 2N ]_{\sqrt{q}} \equiv  - \frac{q^{N}-q^{-N}}
{q^{1/2}-q^{-1/2}} \;.
$$
There exists a matrix representation of $\theta $ and of
$(g,\bar{g})$-differentiation
$\partial$, in which $(E_{+})^{\dag} =E_{-}$ and $N^{\dag} =N$
(or, $\theta^{\dag}=\partial$).
This representation is related to the slightly changed basis for the
algebra $\Gamma_{p}(1)$
$$
\theta^{k} \rightarrow
e^{i\phi_{k}} ( [k]_{\sqrt{q}} !)^{-1/2} \theta^{k} \; ,
$$
where $\phi_{k}$ are arbitrary
real phases.
For each $p$ we obtain different $(p+1)$-dimensional representations
for the algebra $U_{q}(su(1,1))$ when $q$ is a root of unity.
It would be interesting to compare these ``parafermionic representations''
of quantum algebras with other known representations of the similar kind
(see, e.g. [1]).

One might speculate that larger $q$-deformed algebras could be constructed
by virtue of PGA with many $\theta$'s.
Anyway, for further applications one has to develop a detailed theory of PGA
with many variables. In particular, it would allow for a systematic formal
treatment of parasupersymmetries.

As a final remark, we would like to mention a possible relation of PGA to the
finite-dimensional quantum models introduced by H.Weyl in his famous book
[19] and further studied by J.Schwinger (Ref.[20]). They considered quantum
variables described by unitary finite matrices $U_{i}$ satisfying the
relations $U_{i}U_{j}=qU_{j}U_{i}$ and $(U_{i})^{p+1}=1$
(obviously, $q$ must be a root of unity).
They realized that the $p=1$ case is relevant to describing the
spin variables and treated the
infinite-dimensional limit $p \rightarrow \infty $
as a limit in which usual commutative geometry is restored.

\section*{Acknowledgments}
We would like to thank A.B.Govorkov for valuable discussions.


\begin{thebibliography}{99}

 \bibitem{} A.B.Zamolodchikov and V.A.Fateev, \it Sov. Phys. JETP
   \bf 62 \rm(1985) 215; \\
  V.Pasquer and H.Saleur, \it Nucl. Phys. \bf B330\rm(1990)523.

 \bibitem{} R.Mackenzie and F.Wilczek,
   \it Int. J. Mod. Phys. \bf A3 \rm(1988) 2827; \\
  Y.--H.Chen, F.Wilczek, E.Witten and B.Halperin,
   \it Int. J. Mod. Phys. \bf B3 \rm(1989) 1001.

 \bibitem{} C.Aneziris, A.P.Balachandran and D.Sen,
   \it Int. J. Mod. Phys. \bf A6 \rm(1991) 4721; \\
  O.Greenberg, Phys. Rev. \bf D43 \rm(1991) 4111.

 \bibitem{} A.B.Govorkov, \it Sov. J. Part. Nucl. \bf 14 \rm(1983) 520; \\
  Y.Ohnuki and S.Kamefuchi, \it Quantum Field Theory and Parastatistics, \\
	\rm Univ. of Tokyo Press, 1982.

 \bibitem{} V.P.Spiridonov,in: \it Proc. of the Intern. Seminar Quarks-90,
   \rm Telavi, USSR, May 1990; eds. V.A.Matveev et al.,
        World Sci., Singapore, 1990;

 \bibitem{} F.A.Berezin, \it Introduction into Algebra and Analysis with
    Anticommuting Variables, \rm Moscow State University Press, 1983; \\
  R.Casalbuoni, \it Nuovo Cim., \bf A33 \rm(1976) 389.

 \bibitem{} V.A.Rubakov and V.P.Spiridonov, \it Mod. Phys. Lett.
   \bf A3 \rm(1988) 1337.

 \bibitem{} S.Durand, R.Floreanini, M.Mayrand and L.Vinet,
   \it Phys. Lett. \\   \bf 233B \rm(1989) 158.

 \bibitem{} G.Korchemsky, \it Phys. Lett. \bf 267B \rm(1991) 497; \\
   V.D.Gershun and V.I.Tkach, \it Problems of Nuclear Physics and Cosmic
    Rays \rm (Kharkov University Press), \bf 23 \rm(1985) 42.

 \bibitem{} N.Jacobson, \it   Structure of Rings,
   \rm Am. Math. Soc. Publ., Providence, 1956.

 \bibitem{} N.Jacobson, \it   Lie Algebras,
   \rm Interscience Publish, N.-Y.-London, 1962.

 \bibitem{} A.MacFarlane, \it J. Phys. \bf A22 \rm(1989) 4581;  \\
  L.Biedenharn, \it J. Phys. \bf A22 \rm(1989) L873.


 \bibitem{} Yu.I.Manin, \it Quantum Groups and Noncommutative Geometry, \\
   \rm prep. CRM-1561, Montreal, 1988 ; \\
  L.D.Faddeev, N.Yu.Reshetikhin and M.I.Takhtajan, \\
   \it Algebra and Analysis  \bf 1 \rm(1989) 178.

 \bibitem{} E.Corrigan, D.B.Fairlie and P.Fletcher,
   \it J. Math. Phys. \bf 31 \rm(1990) 776.

 \bibitem{} M.Rausch de Traubenberg and N.Fleury,
  in: \it Leite Lopes Festschrift, \\   \rm World Sci., Singapore, 1988.

 \bibitem{} J.Wess and B.Zumino, \it  Nucl. Phys. (Proc. Suppl.)
   \bf 18B \rm(1990) 302.

 \bibitem{} P.P.Kulish and E.V.Damaskinsky,
   \it J. Phys. \bf A23 \rm(1990) L415.

  \bibitem{} M.Chaichian, P.Kulish and J.Lukierski,
    \it Phys. Lett. \bf 262B \rm(1991) 43; \\
   D.B.Fairlie and C.K.Zachos, \it Phys. Lett. \bf 256B \rm(1991) 43.

  \bibitem{} H.Weyl, \it Theory of Groups and Quantum Mechanics,
    \rm Dover, N.Y., 1931.

  \bibitem{} J.Schwinger, \it Proc. Natl. Acad. of Sci., \bf 46 \rm(1960) 570.

\end{thebibliography}
\end{document}